\documentclass{ifacconf}
\usepackage{graphicx}      
\usepackage{natbib}        
\usepackage{amsmath}
\usepackage{subfigure}
\usepackage{comment}
\usepackage{units}
\begin{document}
\begin{frontmatter}

\title{Nonlinear Model Predictive Control for Enhanced Navigation of Autonomous Surface Vessels}  


\author[First]{Daniel Menges} 
\author[First]{Trym Tengesdal} 
\author[First,Second]{Adil Rasheed}

\address[First]{Department of Engineering Cybernetics, Norwegian University of Science and Technology, Trondheim, Norway}
\address[Second]{Department of Mathematics and Cybernetics, SINTEF Digital, Trondheim, Norway}
\address{e-mail: daniel.menges@ntnu.no, trym.tengesdal@ntnu.no, adil.rasheed@ntnu.no}

\begin{abstract}                
This article proposes an approach for collision avoidance, path following, and anti-grounding of autonomous surface vessels under consideration of environmental forces based on Nonlinear Model Predictive Control (NMPC). Artificial Potential Fields (APFs) set the foundation for the cost function of the optimal control problem in terms of collision avoidance and anti-grounding. Depending on the risk of a collision given by the resulting force of the APFs, the controller optimizes regarding an adapted heading and travel speed by additionally following a desired path. For this purpose, nonlinear vessel dynamics are used for the NMPC. To extend the situational awareness concerning environmental disturbances impacted by wind, waves, and sea currents, a nonlinear disturbance observer is coupled to the entire NMPC scheme, allowing for the correction of an incorrect vessel motion due to external forces. In addition, the most essential rules according to the Convention on the International Regulations for Preventing Collisions at Sea (COLREGs) are considered.
The results of the simulations show that the proposed framework can control an autonomous surface vessel under various challenging scenarios, including environmental disturbances, to avoid collisions and follow desired paths. 
\end{abstract}


\end{frontmatter}
{\let\thefootnote\relax\footnotetext{\begin{center}© 2024 the authors. This work has been accepted to IFAC for publication under a Creative Commons Licence CC-BY-NC-ND\end{center}}}

\section{Introduction}
Safe navigation of Autonomous Surface Vessels (ASVs) is a challenging task considering the existence of other dynamic objects, changing weather and sea conditions, and the risk of grounding. In addition, the Convention on the International Regulations for Preventing Collisions at Sea (COLREGs) must be followed according to the \cite{InternationalMaritimeOrganization}, particularly in case of give-way scenarios. 
The state-of-the-art collision avoidance and path planning strategies are elaborately summarized by \cite{vagale_path_2021}, by \cite{huang_ship_2020}, and by \cite{lyu_ship_2023}.

In this regard, several studies have shown that Model Predictive Control (MPC) and its nonlinear version (NMPC) can provide a suitable approach for collision avoidance (COLAV) of ASVs. For instance, \cite{johansen_ship_2016} propose a scenario-based MPC for COLAV considering COLREGs. However, it is mentioned that the existence of environmental disturbances can critically affect the vessel's motion.
The study conducted by \cite{hagen_mpc-based_2018} presents a COLREGs-compliant MPC approach for surface vessels. However, the proposal relies on a kinematical model leading to model simplifications.
In addition, \cite{abdelaal_nonlinear_2018} demonstrate an NMPC-based trajectory tracking and COLAV method defining elliptical ship domains and a nonlinear disturbance observer for improving robustness against the environment. Nevertheless, the NMPC approach relies on other vessels' velocity, course, and length information, which is often unavailable.
The design of an MPC coupled with potential fields has been proven to be a feasible solution regarding other autonomous systems, e.g., autonomous vehicles. For instance, \cite{lin_adaptive_2022} introduces an adaptive potential field for connected autonomous vehicles that incorporates additional motion information (such as heading angle and steering wheel angle) and decouples the APF from the MPC to improve the computation time. In parallel, \cite{elmi_path_2018} propose an optimal path planning technique for autonomous vehicles in dynamic environments, utilizing MPC that decides between lane-keeping and lane-changing maneuvers. By incorporating two distinct potential field functions, one for road boundaries and another for obstacles, the approach ensures vehicular safety.
In \cite{han_potential_2022}, a potential field-based Extended Dynamic Window Approach (EDWA) for automated berthing is proposed, combining global path search, COLREGs-compliant collision avoidance, and linear MPC. However, the method is constrained to mild environmental conditions and unreliable for changing sea conditions.
Since we could not find any literature describing a robust control scheme for COLAV, anti-grounding, and path following under consideration of environmental disturbances, this article tries to tackle these challenges.
Therefore, we propose a robust Nonlinear Model Predictive Control (NMPC) approach for COLAV and path following based on Artificial Potential Fields (APFs) and an environmental disturbance observer.

\section{Preliminaries}
This section covers the necessary foundations used for this study, including the nonlinear vessel dynamics, a nonlinear disturbance observer, the concept of APFs, and the idea of nonlinear MPC. 

\subsection{Vessel Model}
The theory concerning the kinematics and kinetics of a vessel is adopted from \cite{fossen_handbook_2011} and the used model is additionally described in \cite{menges_environmental_2023}. The vessel's kinematical states $\boldsymbol{\eta} = [x_s,  y_s,  \psi]^\top$ express the position regarding the global coordinates $[x, y]^\top$, and the heading $\psi$. Moreover, $\boldsymbol{\nu} = [u,  v,  r]^\top$ contains surge $u$, sway $v$, and yaw rate $r$ leading to
\begin{equation}
	\boldsymbol{\dot{\eta}}=\mathbf{R}_{\mathrm{rot}}(\psi)\boldsymbol{\nu},
\end{equation}
where $\mathbf{R}_{\mathrm{rot}}(\psi)$ defines the rotational matrix, given by
\begin{equation}
	\mathbf{R}_{\mathrm{rot}}(\psi)=\begin{bmatrix}
		\mathrm{cos}(\psi) && -\mathrm{sin}(\psi) && 0\\
		\mathrm{sin}(\psi) && \mathrm{cos}(\psi) && 0 \\
		0 && 0 && 1
	\end{bmatrix}.
\end{equation}
Considering a mass matrix $\mathbf{M}$, a nonlinear damping matrix $\mathbf{D}(\boldsymbol{\nu})$, and a Coriolis and centripetal matrix $\mathbf{C}(\boldsymbol{\nu})$ respectively, the dynamics of a vessel can be expressed by 
\begin{equation}
	\mathbf{M}\boldsymbol{\dot{\nu}}+\mathbf{D}(\boldsymbol{\nu})\boldsymbol{\nu}+\mathbf{C}(\boldsymbol{\nu})\boldsymbol{\nu}=\boldsymbol{\tau}+\boldsymbol{\tau_d}, \label{eq: dynamics}
\end{equation}
where $\boldsymbol{\tau}$ is the control input, and $\boldsymbol{\tau_d}$ are environmental disturbances impacted by wind, waves, and sea currents.
Rewriting the kinematics and kinetics in control-theoretical notation, the state space model is given by
\begin{equation}
	\hspace{-0.6em}\resizebox{.91\hsize}{0.022\vsize}{$\mathbf{\dot{x}} = \begin{bmatrix}
	    \mathbf{0}_{3x3} & \mathbf{R}_{\mathrm{rot}}(\mathbf{x})\\
            \mathbf{0}_{3x3} & \qquad  \mathbf{M}^{-1}[-\mathbf{C}(\mathbf{x})-\mathbf{D}(\mathbf{x})]
	\end{bmatrix}\mathbf{x}+\begin{bmatrix}
	    \mathbf{0}_{3x3}\\
            \mathbf{M}^{-1}
	\end{bmatrix} \mathbf{u}+\begin{bmatrix}
	    \mathbf{0}_{3x3}\\
            \mathbf{M}^{-1}
	\end{bmatrix} \mathbf{d}$}, \label{state_space_model}
\end{equation}
where 
\begin{align}
    \mathbf{x}&=[x_s, y_s, \psi, u, v, r]^\top\\
    \mathbf{u}&= \boldsymbol{\tau} =[\tau_u, 0, \tau_r]^\top\\
    \mathbf{d}&=\boldsymbol{\tau_d}=[\tau_{d,1},\tau_{d,2},\tau_{d,3}]^\top,   
\end{align}
yielding under-actuated vessel dynamics with control inputs $\tau_u$ (thrust in surge) and $\tau_r$ (moment regarding yaw).

\subsection{Nonlinear Disturbance Observer} \label{sec:Disturbance_observer}
The nonlinear disturbance observer developed by \cite{menges_environmental_2023} is used in this work to improve the control behavior obtained by NMPC with extended awareness of the situation. 
The foundation of the observer sets 
\begin{align}
	\boldsymbol{\hat{\tau}_d}&=\boldsymbol{\zeta} + \boldsymbol{\mu}(\boldsymbol{\nu}),\\
	\boldsymbol{\dot{\zeta}}&=\boldsymbol{h}(\boldsymbol{\nu},\boldsymbol{\hat{\tau}_d}),
\end{align}
where the estimation of the disturbances $\boldsymbol{\hat{\tau}_d}$ is defined as the sum of an observer variable $\boldsymbol{\zeta}$ and an unknown mapping $\boldsymbol{\mu}(\boldsymbol{\nu})$.

\cite{menges_environmental_2023} show that the error dynamics of the observer can be obtained to
\begin{align}
	\boldsymbol{\dot{z}} = -\begin{bmatrix}
		\Gamma_1\sigma && 0 && 0\\[6pt]
		0 && \Gamma_2\sigma && 0 \\[6pt]
		0 &&  0 && \Gamma_3\sigma
	\end{bmatrix}\begin{bmatrix}
		z_1\\
		z_2\\
		z_3
	\end{bmatrix}, \label{eq:error_dynamics}
\end{align}
where $\Gamma_1$, $\Gamma_2$, and $\Gamma_3$ are adaptive gains, and
\begin{equation}
    \sigma=1-\frac{\kappa_{23}\kappa_{32}}{\kappa_{22}\kappa_{33}},
\end{equation}
where $\kappa_{ij}$ are the entries of the inverted mass matrix $\mathbf{M}$.
If $\boldsymbol{T}= \frac{\partial \boldsymbol{\mu}(\boldsymbol{\nu})}{\partial \boldsymbol{\nu}}$ is chosen as follows
\begin{equation}
	\boldsymbol{T}=\begin{bmatrix}
		\Gamma_1\frac{1}{\kappa_{11}}\sigma && 0 && 0\\[6pt]
		0 && \Gamma_2\frac{1}{\kappa_{22}} && -\Gamma_2 \frac{\kappa_{23}}{\kappa_{22}\kappa_{33}}\\[6pt]
		0 && -\Gamma_3 \frac{\kappa_{32}}{\kappa_{22}\kappa_{33}} &&\Gamma_3\frac{1}{\kappa_{33}}
	\end{bmatrix},
\end{equation}
globally exponentially stable error dynamics are given by \eqref{eq:error_dynamics}.
The final formulation of the disturbance observer is expressed by
\begin{equation}
	\boldsymbol{\hat{\tau}_d} = \boldsymbol{\zeta}+\boldsymbol{T}\boldsymbol{\nu}, \label{eq:disturbance_observer}
\end{equation}
and
\begin{equation}
	\boldsymbol{\dot{\zeta}} = -\boldsymbol{T}\boldsymbol{\dot{\nu}}(\boldsymbol{\tau_d}=\boldsymbol{\zeta}+\boldsymbol{T}\boldsymbol{\nu}),
\end{equation}
where $\boldsymbol{\hat{\tau}_d}$ is the estimation of the environmental disturbances, and $\boldsymbol{\dot{\zeta}}$ describes the observer update.

\subsection{Artificial Potential Fields}
Artificial Potential Fields (APFs) are commonly used in robotics for navigation and obstacle avoidance. The main idea is to treat a robot as a point charge moving in a potential field, where the goal is an attracting potential and the obstacles introduce repelling potentials.

The total potential $U(\mathbf{q})$ at a point $\mathbf{q}$ in the workspace is the sum of the attractive potential $U_{\text{att}}$ due to the goal and the repulsive potential $U_{\text{rep}}$ due to obstacles yielding

\begin{equation}
U(\mathbf{q}) = U_{\text{att}}(\mathbf{q}) + U_{\text{rep}}(\mathbf{q})
\end{equation}

\subsubsection{Attractive Potential:}
The attractive potential is usually defined as a parabolic pulling force around the goal defined as

\begin{equation}
U_{\text{att}}(\mathbf{q}) = \frac{1}{2} \epsilon ||\mathbf{q} - \mathbf{q}_{\text{goal}}||^2
\end{equation}

where $\epsilon$ is a positive constant, $\mathbf{q}$ is the current position, and $ \mathbf{q}_{\text{goal}}$ is the goal position.

\subsubsection{Repulsive Potential:}
The repulsive potential is introduced by obstacles. For an obstacle $O_i$ at position $ \mathbf{q}_{O_i}$, the repulsive potential is given by

\begin{equation}
	\hspace{-0.8em}\resizebox{.92\hsize}{0.03\vsize}{$U_{\text{rep}}(\mathbf{q}) = 
\begin{cases} 
\frac{1}{2} \eta \left( \frac{1}{||\mathbf{q} - \mathbf{q}_{O_i}||} - \frac{1}{d_{\text{safe}}} \right)^2 & \text{if } ||\mathbf{q} - \mathbf{q}_{O_i}|| < d_{\text{safe}} \\
0 & \text{otherwise}
\end{cases}$} \label{eq:APF_potential}
\end{equation}

where $ \eta$ is a positive constant (repulsive coefficient) and $d_{\text{safe}}$ is a distance threshold within which the obstacle induces a repulsive force. With regard to an ASV, the distance threshold is given by the sensor range of, for instance, Radio Detection and Ranging (RADAR) or Light Detection and Ranging (LiDAR).

\subsubsection{Forces and Motion}
The force acting on the autonomous system due to the potential field is the negative gradient of the total potential yielding

\begin{equation}
\boldsymbol{F}(\mathbf{q}) = -\nabla U(\mathbf{q}).
\end{equation}

For the repulsive potentials, which we later use as COLAV potentials, the force vector is expressed by
\begin{equation}
    \hspace{-0.6em} \resizebox{.90\hsize}{0.02\vsize}{$\boldsymbol{F}_{\text{rep}}(\mathbf{q}) = \eta \left(\frac{1}{||\mathbf{q} - \mathbf{q}_{O_i}||} - \frac{1}{d_{\text{safe}}}\right) \frac{1}{||\mathbf{q} - \mathbf{q}_{O_i}||^3} (\mathbf{q} - \mathbf{q}_{O_i})$}. \label{eq:APF_force}
\end{equation}

As a result, the autonomous system moves in the direction of the force exerted by the potential field, allowing it to navigate toward the goal while avoiding obstacles.

\subsection{Nonlinear Model Predictive Control}

Model Predictive Control (MPC) is a widely used advanced control strategy. In MPC and its nonlinear version (NMPC), the control action is determined by solving an optimization problem at each time step. The optimization aims to minimize a cost function over a prediction horizon, subject to system dynamics and constraints.

\subsubsection{Problem Formulation}

The discretized NMPC problem can be formulated as

\begin{equation}
\hspace{-0.75em}\resizebox{.9\hsize}{0.072\vsize}{$
\begin{aligned}
& \underset{\mathbf{x}(t),\mathbf{u}(t)}{\text{min}}
& & J(\mathbf{x}(t), \mathbf{u}(t)) = \sum_{k=0}^{N-1} l(\mathbf{x}(k|t), \mathbf{u}(k|t)) + V_f(\mathbf{x}(N|t)) \\
& \ \ \ \text{s.t.}
& & \mathbf{x}(0|t) = \mathbf{x}(t), \\
&&& \mathbf{x}(k+1|t) = f(\mathbf{x}(k|t), \mathbf{u}(k|t)), \ \  \forall k \in [0, N - 1] \\
&&& \mathbf{g}_h(\mathbf{x}(k|t), \mathbf{u}(k|t)) \leq 0, \ \  \forall k \in [0, N - 1] \\
&&& \mathbf{g}_s(\mathbf{x}(k|t), \mathbf{u}(k|t)) \leq \mathbf{z}(k|t), \ \  \forall k \in [0, N - 1] \\
&&& \mathbf{z}(k|t) \geq 0, \ \  \forall k \in [0, N - 1] \\
\end{aligned} $}
\end{equation}

Here, $J$ is the cost function over the prediction horizon $N$. The function $l(\mathbf{x}, \mathbf{u})$ represents the stage cost and $V_f(\mathbf{x})$ is the terminal cost. The notation $k|t$ within the brackets indicates a specific time step $k$ associated with the time parameter $t$. It is used to denote the time dependency of the variables. The system dynamics are given by the function $f(\mathbf{x}, \mathbf{u})$. The constraints $\mathbf{g}_h(\mathbf{x}, \mathbf{u}) \leq 0$ represent the hard constraints. The soft constraints are represented by $\mathbf{g}_s(\mathbf{x}, \mathbf{u}) \leq \mathbf{z}$, where $\mathbf{z}$ are slack variables introduced to allow minor violations in constraints. The minimization of these violations can be ensured by penalizing the slack variables in the cost function.

\subsubsection{Hard vs. Soft Constraints}

In NMPC, constraints are either \textit{hard} or \textit{soft}. Hard constraints strictly enforce limits on system states and control inputs and can never be violated. In contrast, soft constraints introduce flexibility by allowing minor violations. This flexibility can be beneficial when dealing with model uncertainties or when a feasible solution respecting hard constraints does not exist. The slack variables associated with soft constraints are usually penalized in the cost function to limit their magnitude.

\section{Methodology} \label{sec:Methodology}
This section covers the original theoretical contributions, including novel concepts for COLAV, anti-grounding, and adapted path following considering disturbances.

\subsection{Collision Avoidance}
The core of the collision avoidance (COLAV) approach is built by APFs. To each tracked target, a repulsive potential field according to \eqref{eq:APF_potential} and \eqref{eq:APF_force} is assigned. Therefore, the resulting force vector of the potential fields is given by
\begin{equation}
    \boldsymbol{F} =\sum_{i=1}^{n}\boldsymbol{F}_i = \begin{bmatrix}
        F_x\\
        F_y
    \end{bmatrix}, \label{eq:force_vector}
\end{equation}
where $n$ denotes the number of tracked targets within the tracking radius.
The magnitude $\mu$ and the angle $\alpha$ of the resulting vector can be obtained by
\begin{align}
    \mu &= ||\boldsymbol{F}||_2 = \sqrt{F_x^2+F_y^2}\\
    \alpha &= \arctan2\left(\frac{F_y}{F_x}\right)
\end{align}

Given the potential field computations, the desired angle for sliding around the potential fields considering a starboard-oriented heading is defined as
\begin{equation}
	\hspace{-0.8em}\resizebox{.92\hsize}{0.03\vsize}{$\psi_{\text{des}} = 
\begin{cases}
\left(1-\frac{d_{\text{min}}}{r_{\text{safety}}}\right) (\alpha - \gamma) + \frac{d_{\text{min}}}{r_{\text{safety}}} \psi_{\text{LOS}}, & \text{if } d_{\text{min}} \leq r_{\text{safety}} \\
\psi_{\text{LOS}}, & \text{otherwise}
\end{cases}$} \label{sliding_angle}
\end{equation}
where $d_{\text{min}}$ is the distance to the closest track, $r_{\text{safety}}$ describes a safety radius within the COLAV algorithm is triggered, and $\psi_{\text{LOS}}$ characterizes the path following heading. The desired heading regarding the path is chosen to be the Line-Of-Sight (LOS) heading, commonly used in guidance systems and autonomous navigation. A description of the LOS guidance applied to ASVs is given in \cite{fossen_line--sight_2003}. The subtracted angle $\gamma$, for our simulations chosen as $\gamma =\frac{\pi}{4}$, guarantees to slide with a safe distance around the obstacles in starboard direction, ensuring the most essential COLREGs. The sliding behavior around the potential fields can be adjusted by varying the angle $\gamma$. 

An additional safety mechanism is developed by defining a similarly projected behavior to the desired surge speed $u_{\text{des}}$ defined as
\begin{equation}
	\hspace{-0.8em}\resizebox{.92\hsize}{0.03\vsize}{$u_{\text{des}} = 
\begin{cases}
\left(1-\frac{d_{\text{min}}}{r_{\text{safety}}}\right) u_{\text{safety}} + \frac{d_{\text{min}}}{r_{\text{safety}}} u_{\text{sp}}, & \text{if } d_{\text{min}} \leq r_{\text{safety}} \\
u_{\text{sp}}, & \text{otherwise}
\end{cases}$} \label{eq:speed_adapt_obstacle}
\end{equation}
where $u_{\text{sp}}$ characterizes the desired speed plan between each waypoint and $u_{\text{safety}}$ describes the desired speed in proximity to an obstacle, in our simulations chosen as $u_{\text{safety}}~=~\frac{u_{\text{sp}}}{2}$. This extension leads to an adapted (slower) speed depending on the distance to the closest obstacle.

\subsection{Anti-Grounding}
Electronic Navigational Charts (ENCs) are used for anti-grounding to compute the closest grounding point with respect to the vessel's draft. That grounding point is then added as a potential field to the force vector given by \eqref{eq:force_vector}. Anti-grounding is only triggered if $d_{\text{min}}$ is related to the anti-grounding potential field.
The anti-grounding method follows the same scheme as the COLAV method disregarding starboard tilting and is defined as
\begin{align}
\psi_{\text{des}} &= \left(1 - \frac{d_{\text{min}}}{r_{\text{grounding}}}\right) \alpha + \frac{d_{\text{min}}}{r_{\text{grounding}}} \psi_{\text{LOS}},
\end{align}
where $r_{\text{grounding}}$ describes the grounding distance, within anti-grounding is triggered.
In parallel, we also added an adapted speed plan given by
\begin{align}
u_{\text{des}} &= \left(1 - \frac{d_{\text{min}}}{r_{\text{grounding}}}\right) u_{\text{grounding}} + \frac{d_{\text{min}}}{r_{\text{grounding}}} u_{\text{sp}}, \label{eq:speed_adapt_grounding}
\end{align}
leading to slow speed in coast proximity and can be adjusted by changing the speed in grounding region $u_{\text{grounding}}$.


\subsection{Optimal Control Problem}
The optimal control problem (OCP) is approached using direct multiple shooting.
Denoting $\dot{\mathbf{x}} = f(\mathbf{x}, \mathbf{u}, \mathbf{d})$ regarding \eqref{state_space_model}, and the state error as
\begin{equation}
    \Tilde{\mathbf{x}} = \begin{bmatrix}
        0 \\ 0 \\ \psi_{\text{des}} - \psi \\ u_{\text{des}} - u \\ 0 \\ r
    \end{bmatrix}
\end{equation}
the optimal control problem is formulated by
\begin{equation}
\hspace{-0.9em}
\resizebox{.9\hsize}{0.1\vsize}{$
\begin{aligned}
& \underset{\mathbf{u}(t), \mathbf{x}(t)}{\text{min}}
& & J(\mathbf{x}(t), \mathbf{u}(t)) = \sum_{k=0}^{N-1} \Tilde{\mathbf{x}}_k^\top \mathbf{Q} \Tilde{\mathbf{x}}_k + \mathbf{u}_k^\top \mathbf{R}\mathbf{u}_k + \boldsymbol{\xi}_k^\top \mathbf{W} \boldsymbol{\xi}_k  \\ \\
& \ \ \ \text{s.t.}
& &  (a) \quad \mathbf{x}_{k+1} = f(\mathbf{x}_k, \mathbf{u}_k, \mathbf{d}_k), \ \ \forall k \in [0, N - 1]\\
& & & (b) \quad \mathbf{x}_0 = \mathbf{x}(t) \\
& & & (c) \quad \mathbf{x}_{lb} \leq \mathbf{x}_k \leq \mathbf{x}_{ub},  \ \ \forall k \in [0, N] \\
& & & (d) \quad \mathbf{u}_{lb} \leq \mathbf{u}_k \leq \mathbf{u}_{ub},  \ \ \forall k \in [0, N - 1]\\
& & & (e) \quad \mathbf{d}_k = \boldsymbol{\hat{\tau}_d},  \ \ \forall k \in [0, N - 1] \\
& & & (f) \quad -\boldsymbol{\xi}_k \leq \Tilde{\mathbf{x}}_k \leq \boldsymbol{\xi}_k,  \ \ \forall k \in [0, N - 1]\\
& & & (g) \quad \boldsymbol{\xi}_k \geq 0, \ \  \forall k \in [0, N - 1]\\ 
\end{aligned} \label{eq:OCP} $}
\end{equation}

Here, $\mathbf{Q}$ is the matrix that weights state errors, $\mathbf{R}$ defines the matrix that weights the use of control inputs, and $\boldsymbol{\xi}$ denotes the vector containing slack variables, weighted by the matrix $\mathbf{W}$. Furthermore, $\mathbf{x}_{lb}$, $\mathbf{x}_{ub}$, $\mathbf{u}_{lb}$, and $\mathbf{u}_{ub}$ are the lower and upper bounds of the states and control inputs, respectively. The constraint $(e)$ in \eqref{eq:OCP} implies that the nonlinear model dynamics of $(a)$ are extended by the estimated disturbances given by the disturbance observer described in Section~\ref{sec:Disturbance_observer}. That allows an adapted correction of the optimal control inputs due to external forces. Note that the estimated disturbances are not predicted over the time horizon but are considered constant for solving the OCP since the observer can only estimate but not predict.


\section{Simulation Setup}
For simulations, the nonlinear model dynamics of the Telemetron vessel are used (see \cite{hagen_mpc-based_2018}). Telemetron was an autonomous research vessel owned by Maritime Robotics. The simulations are conducted utilizing the framework developed by \cite{trym_tengesdal_simulation_2023}, enabling realistic scenarios including real-world sea mapping using ENCs. We chose a region of Møre og Romsdal in Norway for the simulations. In sum, the following five simulation scenarios are visually presented:
\begin{itemize}
    \item [1.] Head-on
    \item [2.] Crossing give-way
    \item [3.] Overtaking
    \item [4.] Anti-grounding
    \item [5.] Crossing give-way including disturbances
\end{itemize}
A further scenario was conducted to demonstrate the observer estimations, the system states, and the distances to dynamic obstacles and grounding hazards under the presence of model uncertainties. For that purpose, scenarios 4 and 5 were combined, and the model parameters used for the NMPC deviated with an uncertainty of 5\% concerning the ground truth.

MPC parameters used for this study were previously tuned and finally chosen to be 
\begin{align}
    \mathbf{Q} &= \mathrm{diag}(0, 0, \mu, \mu, 0, 150)\\
    \mathbf{R} &= \mathrm{diag}(0, 0, 0, 0, 0, 10^{-4})\\
    \mathbf{W} &= \mathrm{diag}(0, 0, 10^3, 10^3, 0, 10^3).
\end{align}
Note that the desired heading and speed in proximity to APFs are weighted by the resulting magnitude $\mu$.
The prediction horizon was set to $N=60$ with a time step of $\Delta t = \unit[0.5]{s}$ since the computation of the OCP took approximately $\unit[0.4]{s}$, leading to real-time applicability. 
Furthermore, the gains of the disturbance observer were chosen to be $\Gamma_1 = \Gamma_2 = \Gamma_3 = 0.2$, and the repulsive coefficient of the APFs was set to $\eta=1$.

\section{Results and Discussion}
\begin{figure}
    \centering
    \includegraphics[width=0.9\linewidth]{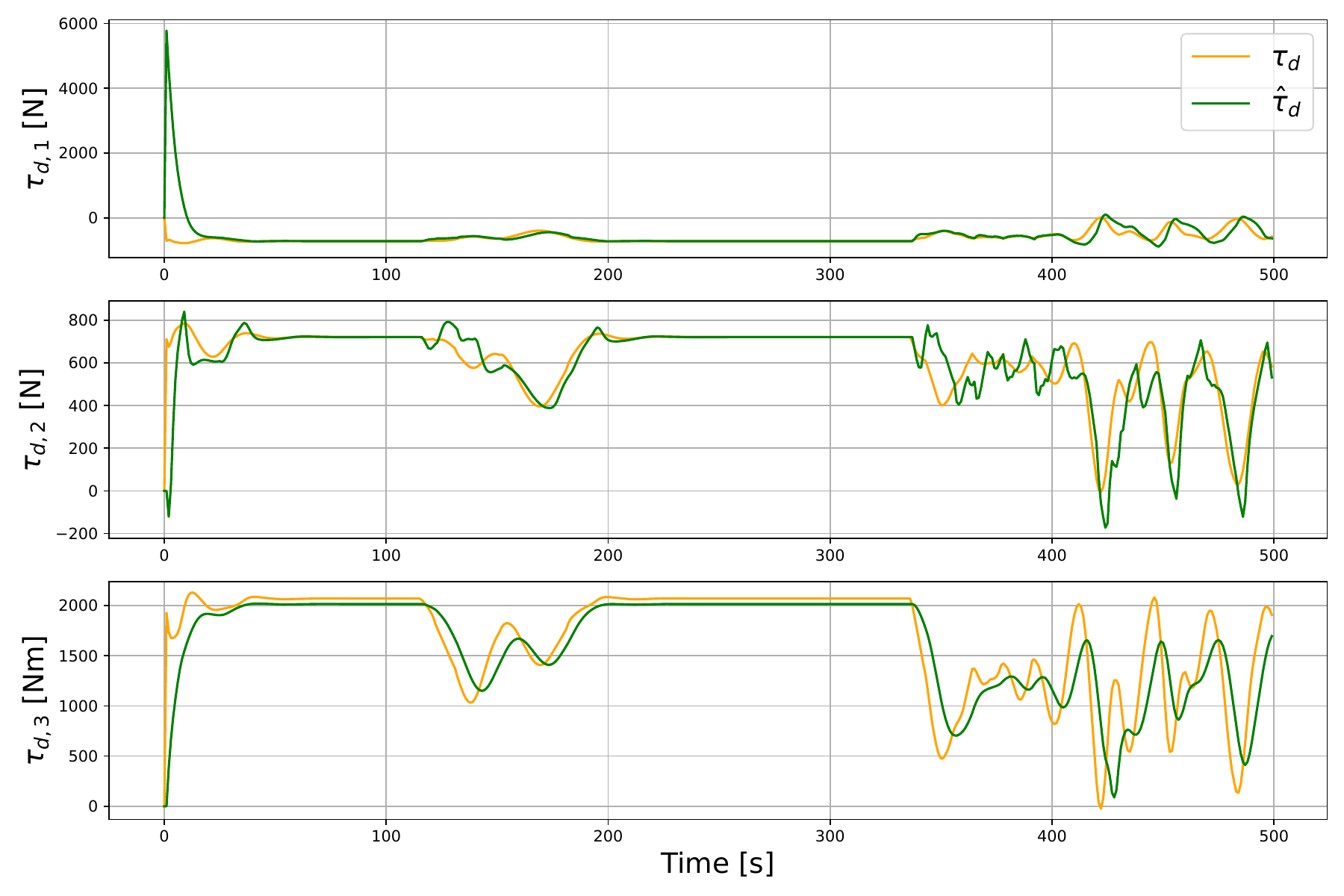}
    \caption{Observed disturbances $\boldsymbol{\hat{\tau}}_d$ in comparison to the simulated disturbances $\boldsymbol{\tau}_d$ with a model parameter deviation of 5\% concerning the ground truth.}
    \label{fig:disturbances}
\end{figure}

It turned out that 2 optimal control inputs per second are sufficient for controlling an ASV since surface vessels are inert systems. 
The estimations of the disturbance observer are shown in Fig.~\ref{fig:disturbances}. Since the time step had to be chosen large, the observer gains were chosen small to ensure stable behavior. Therefore, the adaptation of the observer is slightly delayed.
Considering that the model parameters used for the observer model deviate from the ground truth, the estimations are sufficiently accurate for improving the vessel's behavior via NMPC.

\begin{figure}
    \centering
    \includegraphics[width=0.9\linewidth]{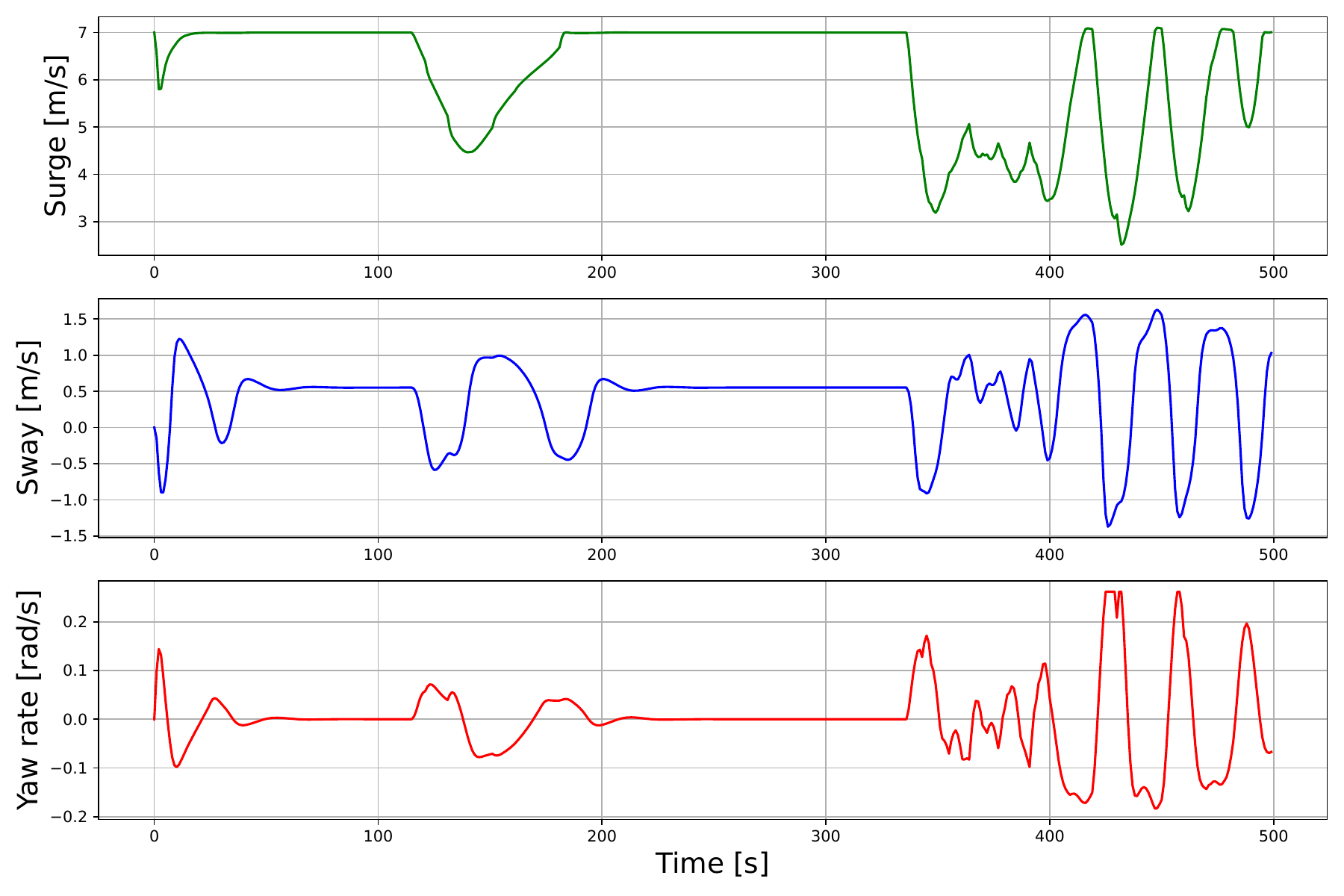}
    \caption{Vessel states (surge, sway, and yaw rate) during a crossing give-way scenario followed by anti-grounding under the impact of environmental disturbances and model uncertainties. The speed plan in surge outside a critical region was set to $\unit[7]{\frac{m}{s}}$. It can be seen how the controller slows down in proximity to a dynamic obstacle $\unit[100]{s}<t<\unit[200]{s}$, and in proximity to grounding hazards $\unit[300]{s}<t<\unit[500]{s}$.}
    \label{fig:states}
\end{figure}

The system states are depicted in Fig.~\ref{fig:states}. The desired speed during the operation was set to $\unit[7]{\frac{m}{s}}$. According to \eqref{eq:speed_adapt_obstacle} and \eqref{eq:speed_adapt_grounding}, the controller adapts the speed of the vessel concerning surge based on the obstacle distance or the grounding distance, respectively. The speed decrease serves as an additional safety mechanism since it enables faster maneuverability due to less inertial drift.

\begin{figure}
    \centering
    \includegraphics[width=0.9\linewidth]{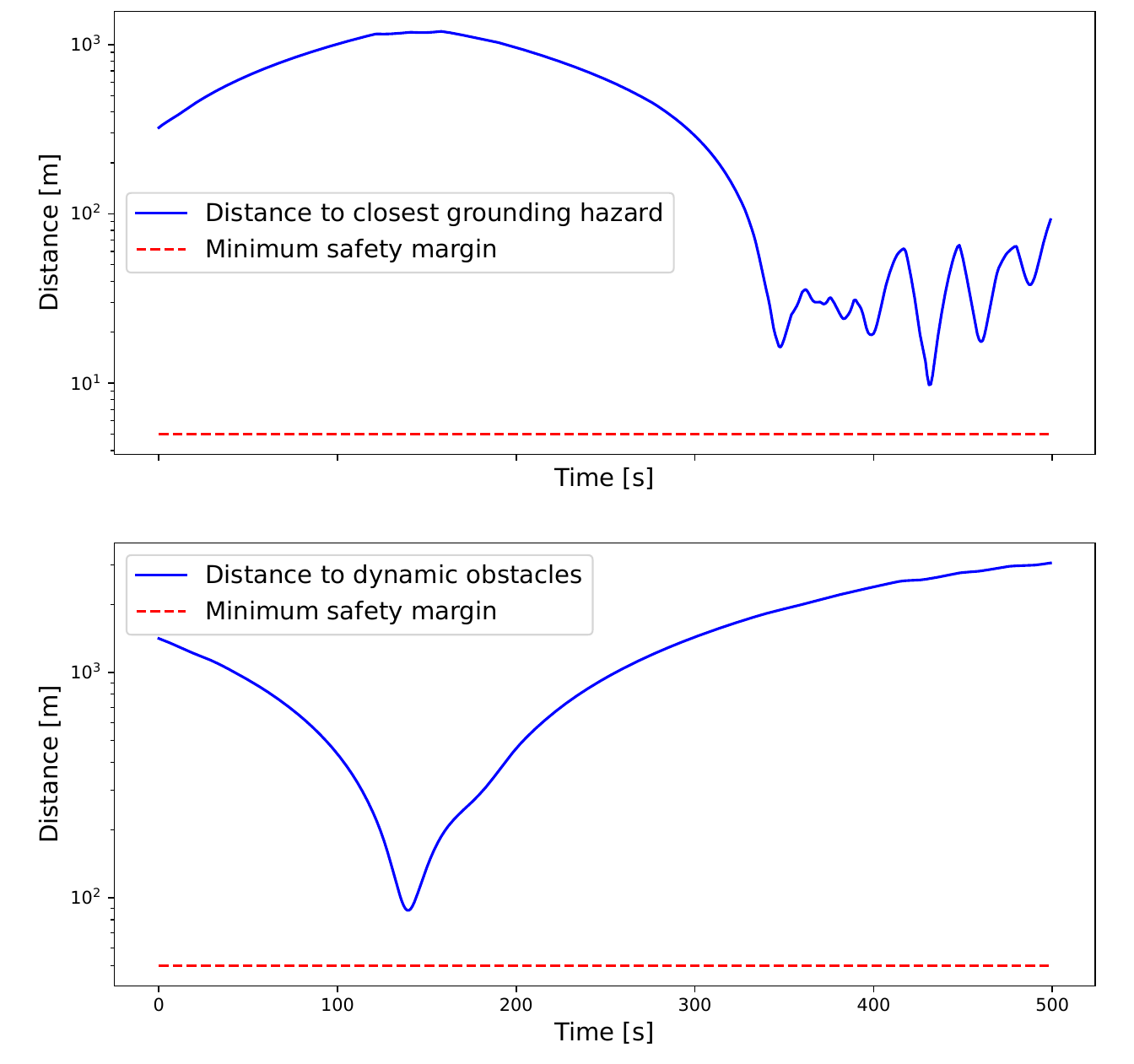}
    \caption{Distance to grounding hazards and dynamic obstacles regarding a crossing give-way scenario under the impact of environmental disturbances.}
    \label{fig:distances}
\end{figure}

Fig.~\ref{fig:distances} presents the distance to the closest dynamic obstacle and the closest grounding hazard. Despite disturbances and model uncertainties, COLAV and anti-grounding is guaranteed. Without the additional speed adaptation (see Fig.~\ref{fig:states}), the minimum safety margin could be exceeded in some cases.

\begin{figure*}[h]
    \centering
    \begin{minipage}{0.3\linewidth}
        \centering
        \includegraphics[width=\linewidth]{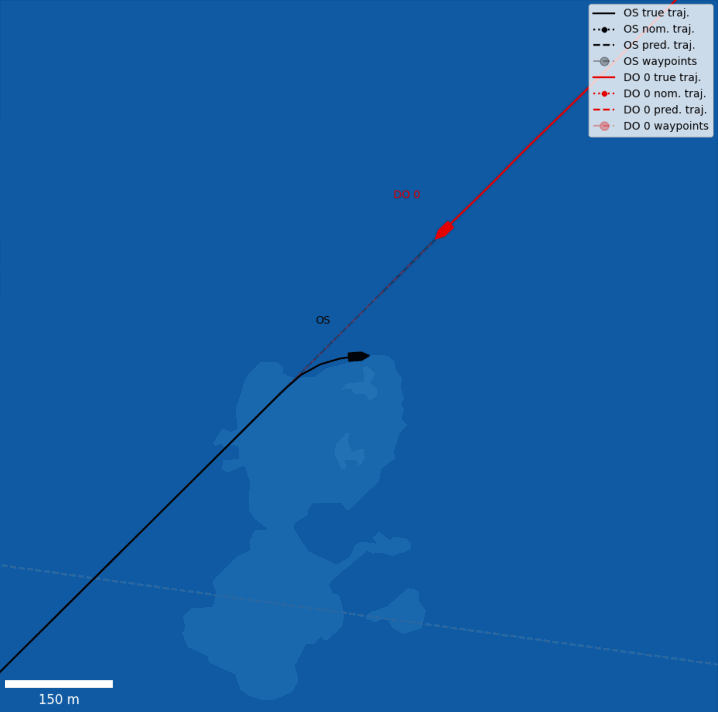}
        \\[1ex] 
    \end{minipage}
    \begin{minipage}{0.3\linewidth}
        \centering
        \includegraphics[width=\linewidth]{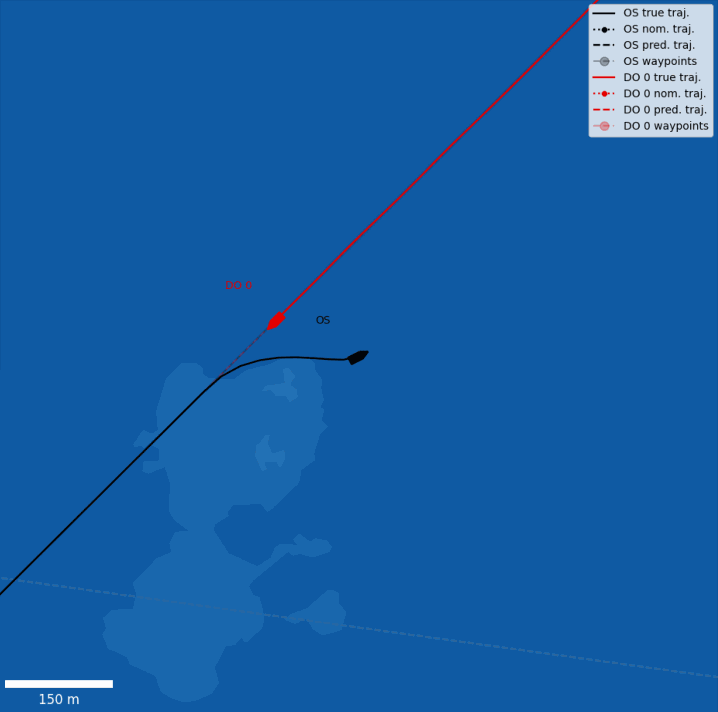}
        \\[1ex] 
    \end{minipage}
    \begin{minipage}{0.3\linewidth}
        \centering
        \includegraphics[width=\linewidth]{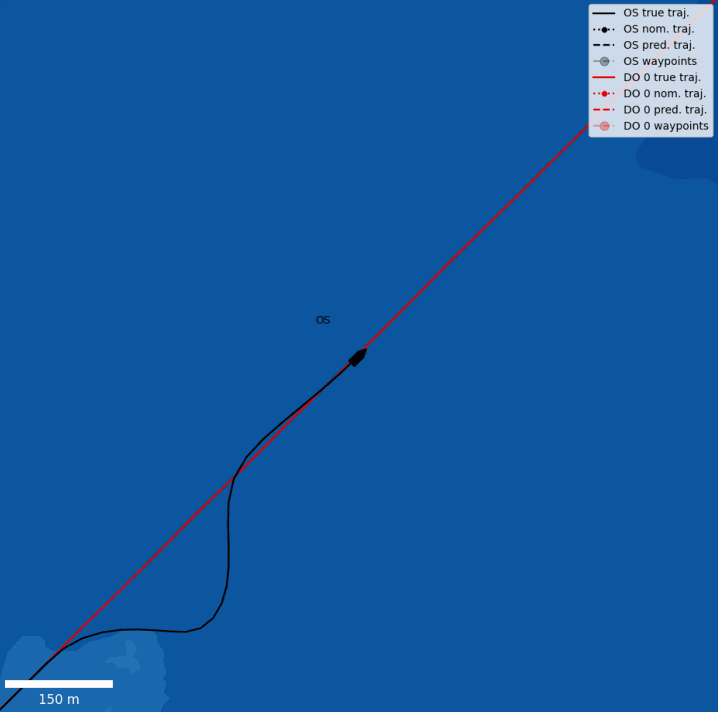}
        \\[1ex] 
    \end{minipage}\\[1ex] 
    Head-on scenario.
    \\[2ex]

    \centering
    \begin{minipage}{0.3\linewidth}
        \centering
        \includegraphics[width=\linewidth]{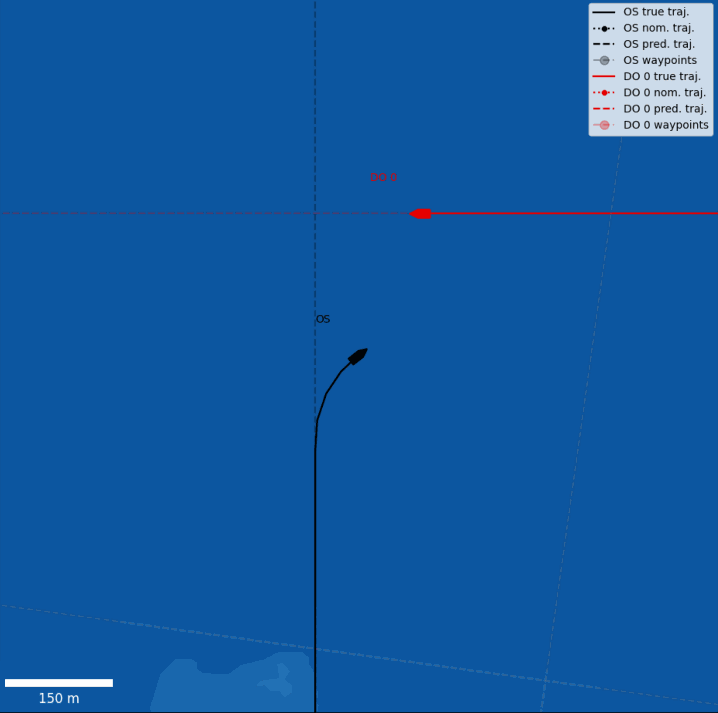}
        \\[1ex] 
    \end{minipage}
    \begin{minipage}{0.3\linewidth}
        \centering
        \includegraphics[width=\linewidth]{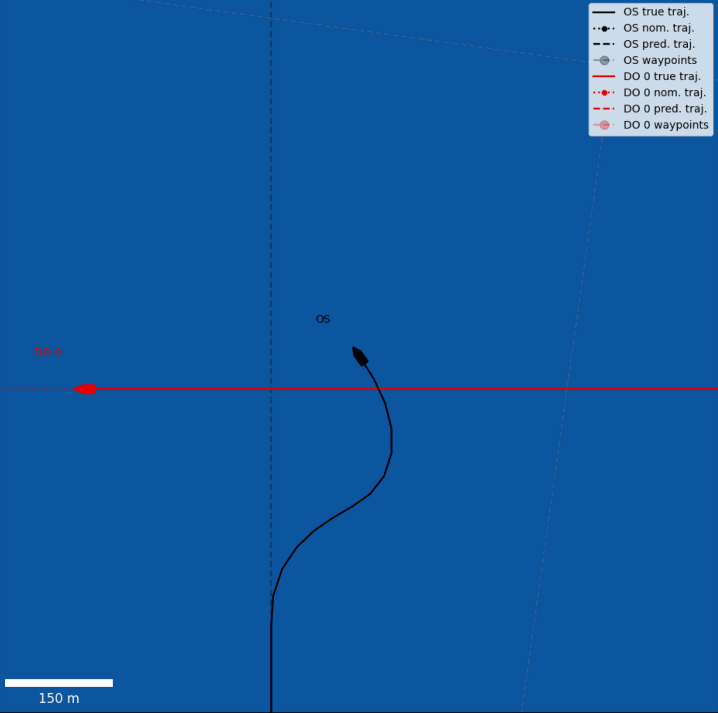}
        \\[1ex] 
    \end{minipage}
    \begin{minipage}{0.3\linewidth}
        \centering
        \includegraphics[width=\linewidth]{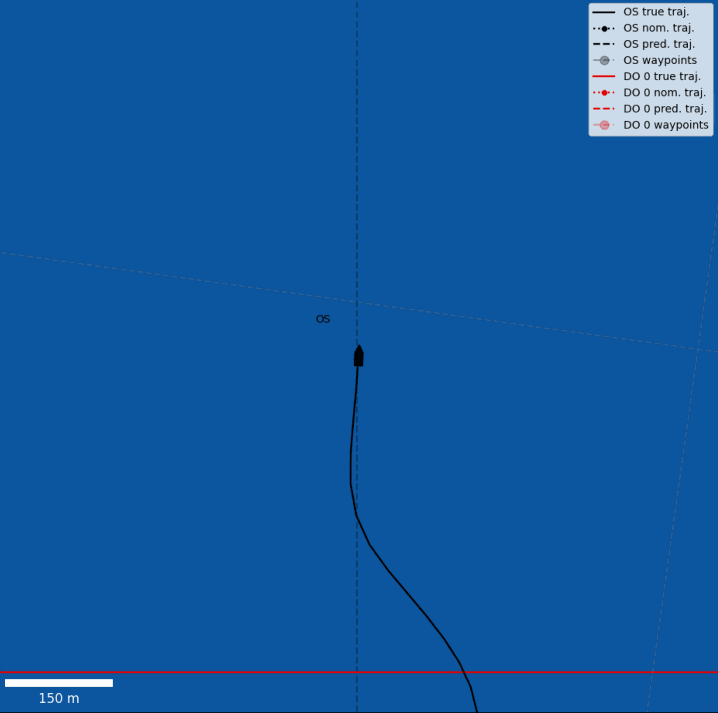}
        \\[1ex] 
    \end{minipage}\\[1ex] 
    Crossing give-way scenario.
    \\[2ex]

    \centering 
    \begin{minipage}{0.3\linewidth}
        \centering
        \includegraphics[width=\linewidth]{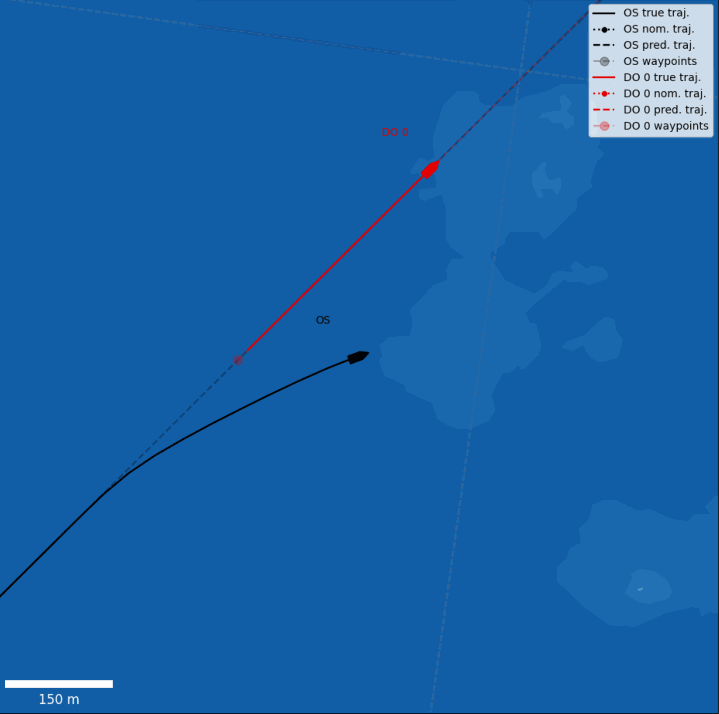}
        \\[1ex] 
    \end{minipage}
    \begin{minipage}{0.3\linewidth}
        \centering
        \includegraphics[width=\linewidth]{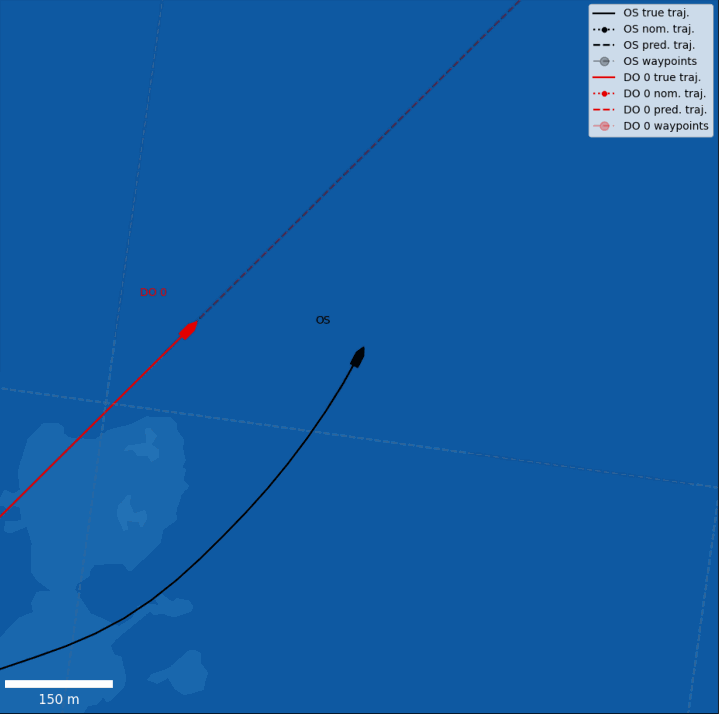}
        \\[1ex] 
    \end{minipage}
    \begin{minipage}{0.3\linewidth}
        \centering
        \includegraphics[width=\linewidth]{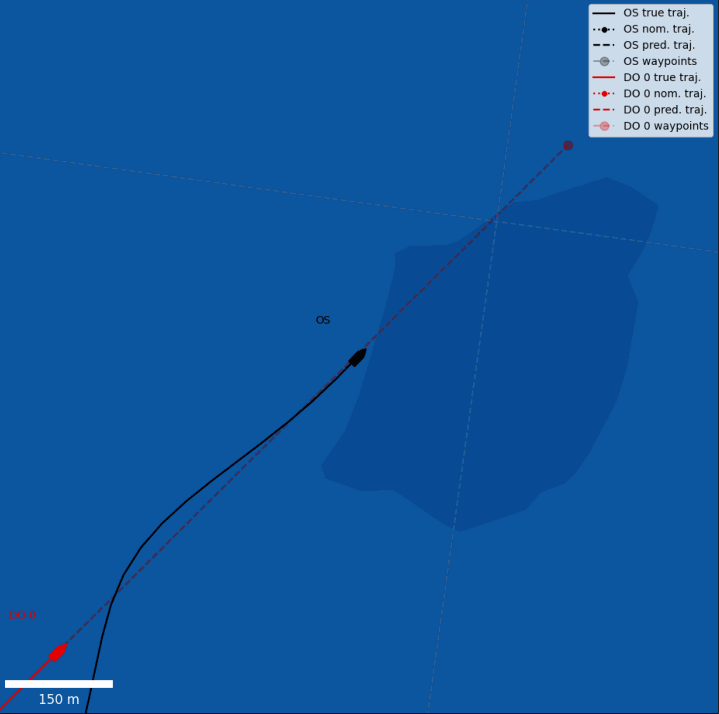}
        \\[1ex] 
    \end{minipage}\\[1ex] 
    Overtaking scenario.
    
    \caption{Collision avoidance maneuvers according to the COLREGS. The controlled ownership (OS) is depicted in black, while the dynamic obstacle (DO) is visualized in red. Presented are, in each case, three snapshots of the scenario in sequential order. In all three cases, the OS successfully initiates a starboard-sided obstacle avoidance maneuver and subsequently follows the desired path.}
    \label{fig:COLREG_scenarios}
\end{figure*}

The COLREGs-compliant scenarios head-on, crossing give-way, and overtaking are shown in Fig.~\ref{fig:COLREG_scenarios}. 
Presented are image sequences of the related scenarios within the used simulator. It is shown that the proposed control scheme consisting of APFs and NMPC can maneuver the ASV in critical situations according to the COLREGs. As described in Section~\ref{sec:Methodology}, tuning parameters such as $\gamma$, $r_{\text{safety}}$, and $u_{\text{safety}}$ can be used to change the desired behavior for COLAV scenarios. That leads to a further opportunity for safety tuning since the OCP optimizes for heading $\psi_{\text{des}}$ and additionally for an adapted surge speed $u_{\text{des}}$. Furthermore, the repulsive coefficient $\eta$ of the potential fields can be adjusted to change the COLAV behavior. If the user increases $\eta$, the proximity of the closest obstacle has a stronger influence regarding the vessel's bearing.

\begin{figure*}
    \centering
    \begin{minipage}{0.3\linewidth}
        \centering
        \includegraphics[width=\linewidth]{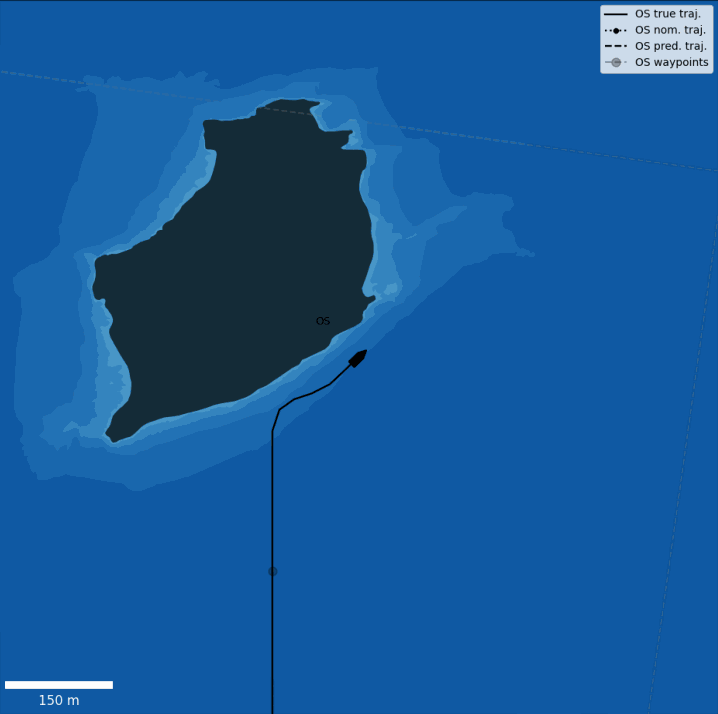}
        \\[1ex] 
    \end{minipage}
    \begin{minipage}{0.3\linewidth}
        \centering
        \includegraphics[width=\linewidth]{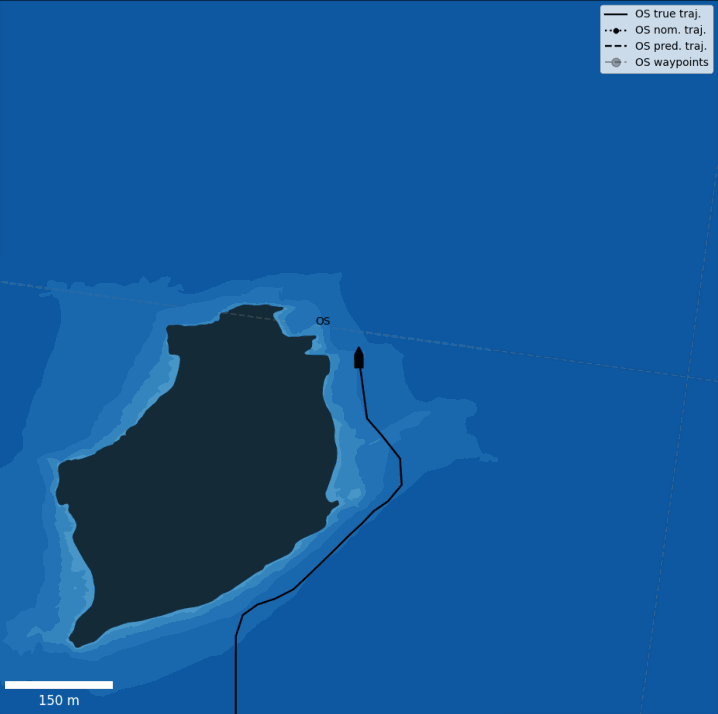}
        \\[1ex] 
    \end{minipage}
    \begin{minipage}{0.3\linewidth}
        \centering
        \includegraphics[width=\linewidth]{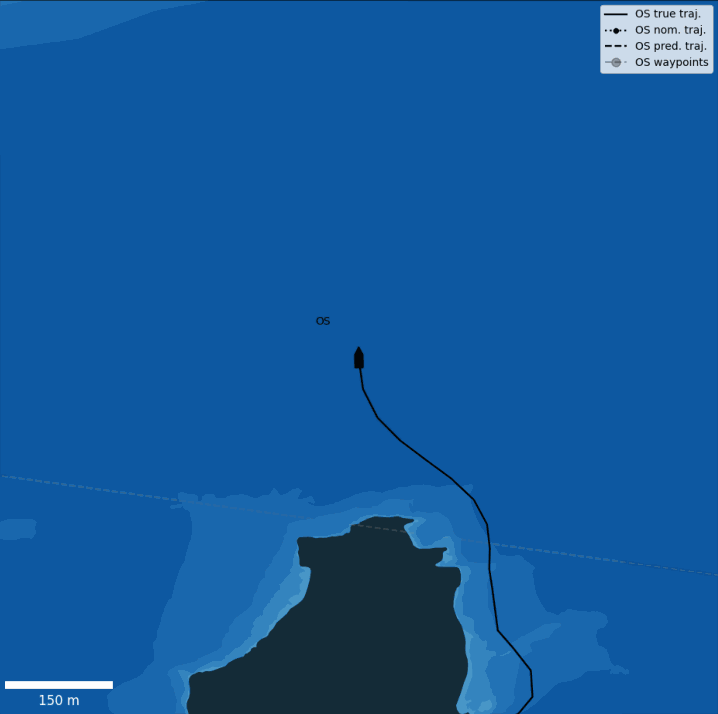}
        \\[1ex] 
    \end{minipage}\\[1ex] 
    \caption{Anti-grounding scenario. Presented are three snapshots of the scenario in sequential order.}
    \label{fig:grounding_scenario}
\end{figure*}

Anti-grounding capabilities are presented in Fig.~\ref{fig:grounding_scenario}, showing the ASV crossing an isle at a safe distance regarding its draft. 
The ability for anti-grounding can be highly relevant for real-world applications since the marine bottom is cluttered by irregularities and sandbanks.

\begin{figure*}
    \centering
    \begin{minipage}{0.3\linewidth}
        \centering
        \includegraphics[width=\linewidth]{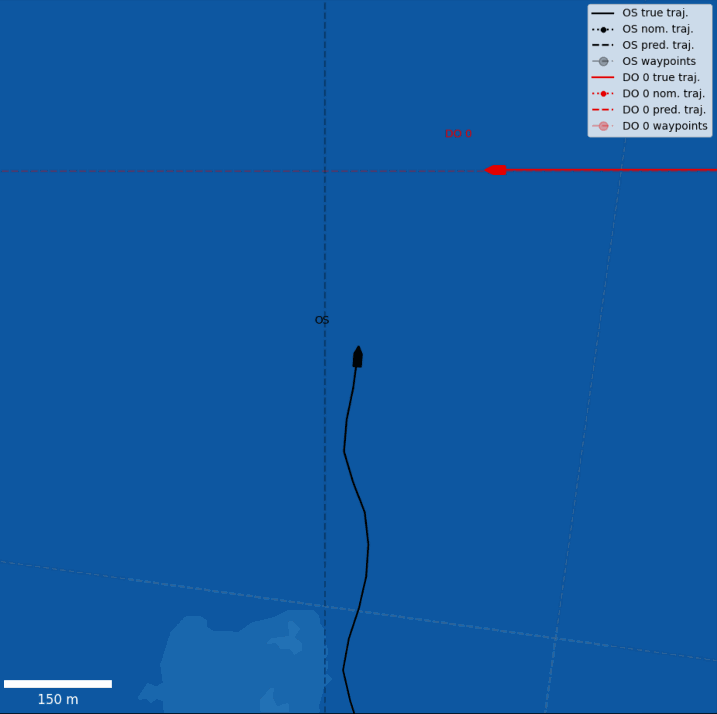}
        (a)
        \\[1ex] 
    \end{minipage}
    \begin{minipage}{0.3\linewidth}
        \centering
        \includegraphics[width=\linewidth]{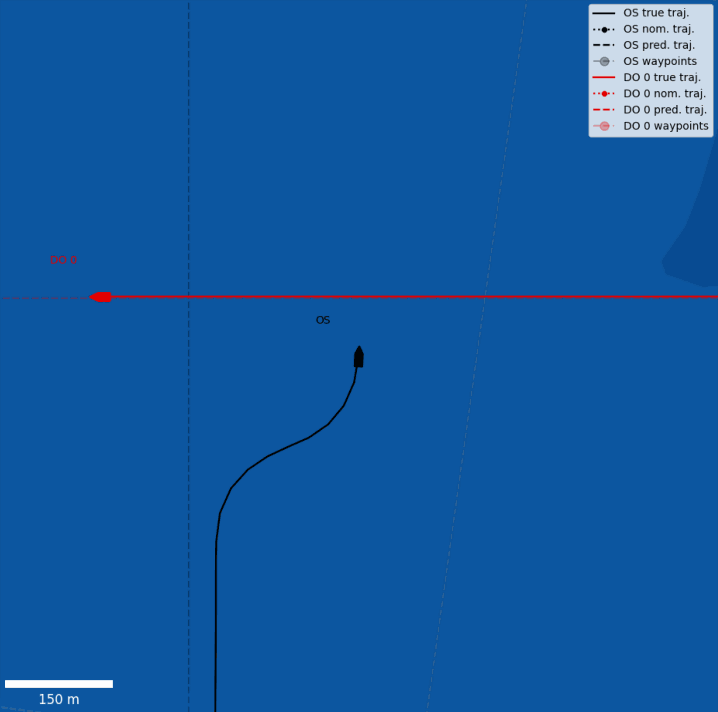}
        (b)
        \\[1ex] 
    \end{minipage}
    \begin{minipage}{0.3\linewidth}
        \centering
        \includegraphics[width=\linewidth]{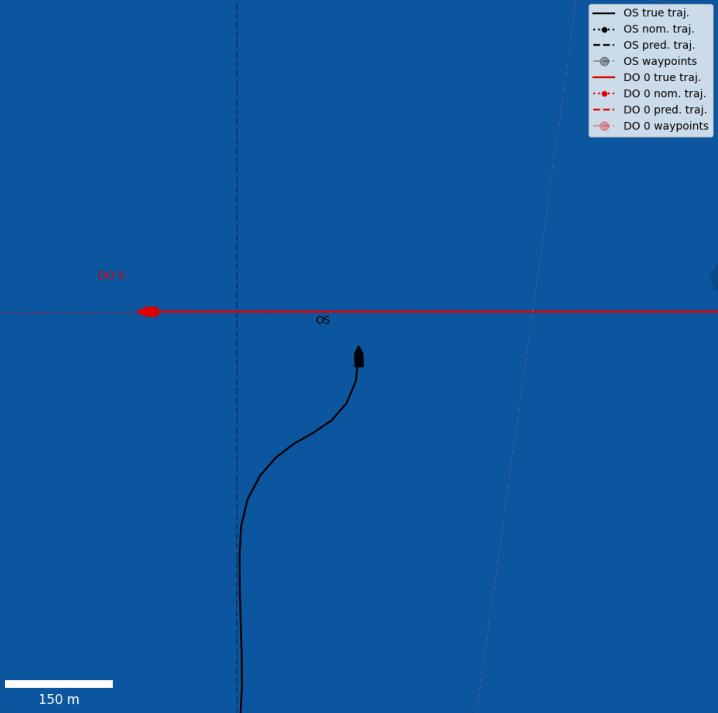}
        (c)
        \\[1ex] 
    \end{minipage}
    \caption{Disturbance correction. The disturbances impact the vessel port-sided. Compared are a PID controller with LOS guidance (a), the raw NMPC scheme without observed disturbances (b), and the NMPC with instilled knowledge of the observed disturbances (c).}
    \label{fig:disturbance_correction}
\end{figure*}

The correction of environmental disturbances is depicted in Fig.~\ref{fig:disturbance_correction}. To demonstrate the improvement using the disturbance observer from Section~\ref{sec:Disturbance_observer}, the proposed approach is compared to a common Proportional–Integral–Derivative (PID) controller, and an NMPC-driven approach disregarding the disturbances $\mathbf{d}$ within $(a)$ and $(e)$ in the OCP formulation described by \eqref{eq:OCP}. Note that the PID controller was tuned such that it worked properly for path following if no disturbances were present. The results show that path following can be adapted to environmental disturbances, guaranteeing safer behavior. While the PID controller leads to strong deviations of the desired path, the NMPC-driven approach disregarding disturbances has likewise a path offset. However, the observer-based NMPC method properly corrects the impact of environmental disturbances. In addition, it was noticed that the disturbance observer, coupled with NMPC, enables improved problem-solving capabilities of the OCP. Without the estimations of the disturbances, the solver struggled to find a solution if the disturbances became more powerful.

The generation of desperate scenarios under the impact of disturbances in which multiple dynamic obstacles approached the vessel close to grounding hazards revealed the limitation of the proposed NMPC. Such extreme scenarios can lead to unexpected rotational behavior of the vessel due to underactuated properties and the rapidly changing desired heading obtained from the resulting APF force.

This study only uses positional estimations of other dynamic obstacles since predictions are often not accessible. However, incorporating the predicted paths of other obstacles into the OCP could improve the behavior of the NMPC approach. As a result, extreme scenarios as described above (multiple dynamic obstacles close to grounding hazards) could be handled without unnecessary movements due to additional proactive situational awareness.


\section{Conclusion}

In summary, this article introduces an innovative approach to collision avoidance and path following for Autonomous Surface Vessels (ASVs) using Nonlinear Model Predictive Control (NMPC) with Artificial Potential Fields (APFs). Integrating APFs ensures effective collision avoidance, anti-grounding, and COLREG compliance while sustaining the desired path.
The significance of this work lies in the capability of ASVs to navigate in complex scenarios, including environmental disturbances, by extending the NMPC problem by estimations obtained by a nonlinear disturbance observer. Therefore, the proposed framework provides a promising approach for real-world applications.
Future work might additionally incorporate predictions of dynamic obstacles and address the optimization of fuel consumption. 

\begin{ack}
This work is part of SFI AutoShip, an 8-year research-based innovation center. 
In addition, this research project is integrated into the PERSEUS doctoral program. 
We want to thank our partners, including the Research Council of Norway, under project number 309230, and the European Union’s Horizon 2020 research and innovation program under the Marie Skłodowska-Curie grant agreement number 101034240.
\end{ack}

\bibliography{references}

\end{document}